\documentclass[letterpaper,twocolumn,10pt]{article}

\usepackage{hyperref} 
\usepackage{usenix-2020-09} 
\usepackage{subcaption} 
\usepackage{booktabs} 
\usepackage{tikz} 
\usepackage{amsmath} 
\usepackage[scaled=0.8]{beramono} 
\usepackage{listings} 
\usepackage{balance} 
\usepackage{pifont} 
\usepackage[maxbibnames=4, sorting=none, backref=true]{biblatex} 
\usepackage{makecell} 

\newcommand{\papername}{Nitriding: A tool kit for building scalable, networked, secure enclaves}
\newcommand{\tool}{nitriding}
\newcommand{\Tool}{Nitriding}
\usetikzlibrary[shapes,arrows,positioning,arrows.meta,calc,fit]

\hypersetup{
  pdftitle={\papername{}},
  pdfauthor={Philipp Winter, Ralph Giles, Moritz Schafhuber, and Hamed Haddadi},
  pdfkeywords={security, privacy, networking, enclave},
  colorlinks=true,
  urlcolor=gray,
  linkcolor=gray,
  citecolor=gray,
}

\begin{document}

\title{\Large \bf \papername{}}

\author{
{\rm Philipp Winter}\\
Brave Software
\and
{\rm Ralph Giles}\\
Brave Software
\and
{\rm Moritz Schafhuber}\\
Brave Software
\and
{\rm Hamed Haddadi}\\
Brave Software \\ Imperial College London
} 

\maketitle

\begin{abstract}

Enclave deployments often fail to simultaneously be
\emph{secure} (e.g., resistant to side channel attacks),
\emph{powerful} (i.e., as fast as an off-the-shelf server), and
\emph{flexible} (i.e., unconstrained by development hurdles).
In this paper, we present \tool{}, an open tool kit that enables the
development of enclave applications that satisfy all three properties.
We build \tool{} on top of the recently-proposed AWS Nitro Enclaves whose
architecture prevents side channel attacks by design, making \tool{} more
secure than comparable frameworks.  We abstract away the constrained
development model of Nitro Enclaves, making it possible to run unmodified
applications inside an enclave that have seamless and secure Internet
connectivity, all while making our code user-verifiable.
To demonstrate \tool{}'s flexibility, we design three enclave applications, each
a research contribution in its own right:
(i) we run a Tor bridge inside an enclave, making it resistant to protocol-level
deanonymization attacks;
(ii) we built a service for securely revealing infrastructure configuration,
empowering users to verify privacy promises like the discarding of IP addresses
at the edge;
(iii) and we move a Chromium browser into an enclave, thereby isolating its
attack surface from the user's system.
We find that \tool{} enables rapid prototyping and alleviates the deployment of
production-quality systems, paving the way toward usable and secure enclaves.

\end{abstract}

\section{Introduction}

Real-world enclave deployments tend to fall short in either providing security
(e.g., resisting side channel attacks), flexibility (i.e., supporting diverse
use cases and being easy to develop), or powerfulness (i.e., enabling
computationally demanding applications).  While reasonably flexible and
powerful, Intel's SGX failed to be secure because of its susceptibility to side
channel attacks~\cite{VanBulck2018a,Murdock2020a,Brasser2017a}.  Newer
developments like Apple's Secure Enclave appear secure but is limited to cell
phones and therefore lacks powerfulness and flexibility.

In this work, we propose \tool{}, a free software tool kit that satisfies the
aforementioned properties.  By building on top of the recently proposed Nitro
Enclaves~\cite{nitro-enclaves}, \tool{} inherits its strong security properties.
Unlike Intel's SGX, Nitro Enclaves run on dedicated CPU and memory that is not
shared with untrusted code, which promises to mitigate side channel attacks.
By default, Nitro Enclaves are however severely constrained and difficult to
develop applications for because (i)~Nitro Enclaves are not meant to run
networked applications; (ii)~remote attestation is not designed to involve third
parties; (iii)~and enclave features like horizontal scaling require proprietary
Amazon technology whose privacy promises cannot be verified.  Our work abstracts
away these constraints, to make \tool{} flexible and user-verifiable.  Our tool
kit provides enclave applications with seamless networking via a tunneled
network interface, and it facilitates the creation of secure and authenticated
channels based on HTTPS.  We also design a new mechanism for the horizontal
scaling of enclaves, all while empowering third parties (i.e., users) to audit
these features using remote attestation.

To demonstrate \tool{}'s usefulness, we put it to the test by developing three
new enclave application, each a research contribution in its own right.
First, we build an application that allows a service provider to ``publish'' its
infrastructure configuration, allowing the provider's users to verify that
privacy promises (e.g., the stripping of IP addresses) are properly configured.
Second, we show how a Tor bridge\footnote{A Tor bridge is a ``private'' Tor
relay whose purpose is to help users circumvent Internet censorship.}
can be run inside an enclave, which protects its users from protocol-level
attacks such as the infamous 2015 attack run by CMU~\cite{Dingledine2015a}.
We found that this enclave-enabled Tor bridge allows for convenient Web
browsing---4K YouTube videos played smoothly and without buffering.
Finally, to show that \tool{} can sustain computationally demanding low-latency
use cases, we launch a Chromium browser inside an enclave and make it available
to users via a remote desktop interface.  This protects users from browser
exploits by constraining the browser to a virtual machine that's physically
separate from the user's machine.

Summing up, this work makes the following contributions:

\begin{itemize}

  \item We design and implement \emph{\tool{}}, a free software enclave tool kit
    that enables the rapid development of secure and computationally demanding
    enclave applications.

  \item We evaluate the latency and throughput guarantees of both \tool{} and
    the underlying Nitro Enclaves, finding that low-latency and high-throughput
    applications are possible despite there being room for much improvement.

  \item We put \tool{} to the test by developing three new enclave applications,
    finding that these applications can comfortably support computationally
    demanding, low-latency, and high-throughput use cases, all while enabling
    rapid prototyping.

\end{itemize}

Next, we provide background (\S~\ref{sec:background}), we present \tool{}'s design
(\S~\ref{sec:design}), we propose three enclave applications built on top of
\tool{} (\S~\ref{sec:applications}), we evaluate \tool{}'s latency and
throughput (\S~\ref{sec:evaluation}), we discuss its limitations
(\S~\ref{sec:limitations}), we discuss related work (\S~\ref{sec:related-work}),
and we finally conclude this paper (\S~\ref{sec:conclusion}).

\section{Background}%
\label{sec:background}

Secure enclaves can take many shapes and offer various security properties, but
in this work we only require the following three.  In the rest of this section,
we provide background on how Nitro Enclaves achieve the above three security
properties.

\noindent \textbf{Confidentiality:} An unauthorized entity (e.g., the host
operating system) must not be able to observe the data that an enclave is
processing.

\noindent \textbf{Integrity:} An unauthorized entity must not be able to modify
the data that the enclave is processing, or the code that it is running.

\noindent \textbf{Verifiability:} Any entity (e.g., a user) must be able to
verify if the enclave is running the code that its operator claims it is
running.

\subsection{The AWS Nitro system}%
\label{sec:nitro}

Nitro Enclaves are virtual machines that run on dedicated hardware that is not
shared with an enclave's EC2 host.  The technology that enforces isolation
between the enclave and its EC2 host also enforces isolation between any two
given EC2 instances: the Nitro system.  Before covering enclaves, we explain how
the Nitro system works---first by discussing its three key components.

\textbf{Nitro cards}: While physically connected to a server's main board via
PCIe, Nitro cards are dedicated and custom-built hardware and software that runs
independently of a server's main board.  Nitro cards implement the interfaces
that allow for the management of a server's computational, memory, and storage
needs, among other things.  A Nitro card also provides a server's hardware root
of trust and is responsible for firmware updates, secure boot, and acts as an
interface between the server and the EC2 control plane~\cite[pp.
7--10]{Bean2022a}.

\textbf{Nitro security chip}: The Nitro card acts independently of the system main
board.  The purpose of the Nitro security chip, which is controlled by the Nitro
card, is to extend the Nitro card's control over the system main board.  One of
the chip's responsibilities is to prevent the CPU from updating the system's
firmware when run in bare metal mode~\cite[pp.~10--11]{Bean2022a}.

\textbf{Nitro hypervisor}:
The hypervisor is a firmware-like component that receives commands from the
Nitro card.  The hypervisor is stripped of any non-essential code: it does not
contain networking code, file systems, shells, or other utilities that would
allow a successful attacker to access other
infrastructure~\cite[pp.~11--12]{Bean2022a}.

Other design decisions are meant to provide defense in depth.  First, by
design, the Nitro system has no operator access, i.e., operators are unable to
log in to an EC2 Nitro system and inspect memory or access customer
data~\cite[p.~15]{Bean2022a}. Second, the Nitro system is designed to
communicate passively, i.e., system components never initiate outgoing
connections during production operations.

Of particular interest is how the Nitro system aims to prevent side channel
attacks: customer instances never share a given CPU core in parallel.  If two
customers use a CPU core sequentially, the hypervisor ensures that state is
cleared in between use.  Depending on the instance, cores may be exclusively
allocated to a customer, which includes Nitro Enclaves.  This means that L1 and
L2 caches are also never shared.  Last-level cache lines may be shared but only
non-simultaneously.  Amazon's documentation further
states~\cite[p.~19]{Bean2022a}:

\begin{quote}
By virtue of its function, only relatively infrequently accessed data is
referenced in last-level cache lines.  Side-channels typically require a very
large and statistically relevant number of samples in order to over-come the
noise present in systems.
\end{quote}

\subsection{Nitro Enclaves}%
\label{sec:nitro-enclaves}

Nitro Enclaves inherit the isolation and security properties of the Nitro
system.  When an EC2 host system launches a Nitro Enclave, it ``sacrifices'' at
least one of its CPUs and some of its memory pages to the enclave.  These
resources are subsequently unavailable to the EC2 host and exclusively used by
the enclave.  The same isolation mechanism that protects individual customer EC2
instances from each other also protects the Nitro Enclave from its host.

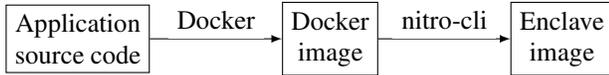
\begin{figure}[t]
  \centering
  \begin{tikzpicture}[node distance=20pt]

  \node [draw,
         align=center] (code) {Application\\source code};

  \node [draw,
         align=center,
         right=50pt of code] (docker) {Docker\\image};

  \node [draw,
         align=center,
         right=50pt of docker] (eif) {Enclave\\image};

  \draw[-latex] (code.east) -- (docker.west)
                node [midway, above, fill=white] {Docker};
  \draw[-latex] (docker.east) -- (eif.west)
                node [midway, above, fill=white] {nitro-cli};

\end{tikzpicture}
  \caption{The development workflow for compiling enclave applications.}%
  \label{fig:dev-workflow}
\end{figure}

On the software level, Nitro Enclaves are virtual machines.  They have their own
Linux kernel that is independent from the host.  Customers can create enclave
images from a Docker image that contains the enclave application.  Amazon
provides a command line tool, nitro-cli~\cite{nitro-cli}, which compiles a
Docker image into an enclave image file (EIF).  Figure~\ref{fig:dev-workflow}
illustrates the process.  After compilation, nitro-cli prints a number of
platform configuration registers (PCRs) that contain SHA-384 hashes over
different layers of the enclave image file.  Table~\ref{tab:pcr} shows the six
available PCRs.  PCR0 is of particular importance for remote attestation as we
will explain later.

\begin{table}[t]
    \centering
    \begin{tabular}{r l}
    \toprule
      PCR \# & SHA-384 hash of\ldots \\
    \midrule
      0 & Enclave image file \\
      1 & Linux kernel \\
      2 & Application \\
      3 & IAM role assigned to the host instance \\
      4 & Instance ID of the host instance \\
      8 & Enclave image file signing certificate \\
    \bottomrule
    \end{tabular}
    \caption{The available platform configuration registers (PCRs) and the
    meaning behind them.}%
    \label{tab:pcr}
\end{table}

By design, Nitro Enclaves have very limited abilities to communicate with the
outside world.  Lacking a dedicated networking interface, Nitro Enclaves can
only communicate with their EC2 host via a VSOCK interface~\cite{vsock}.
Originally proposed for communication between a hypervisor and its virtual
machines, AWS repurposed the VSOCK interface to serve as communication channel
between an enclave and its parent EC2 instance.  From a developer's point of
view, the VSOCK interface is a point-to-point interface connecting the two.  On
the network layer, 32-bit context IDs take the role of IP addresses in VSOCK
interfaces.  For example, the enclave may have context ID 4 while its parent EC2
instance may have context ID 3.  On the transport layer, one can use the same
protocols that one would use over the IP-based address family; namely TCP, UDP,
et cetera.

\section{\Tool{}'s design}%
\label{sec:design}

We now discuss \tool{}'s design by first laying out our trust assumptions and by
providing an informal design overview (\S~\ref{sec:assumptions-overview}),
followed by a discussion of the two major aspects of our tool kit: the
reproducible build system (\S~\ref{sec:build-system}) and \tool{}
itself~(\S~\ref{sec:framework}).

\subsection{Trust assumptions and design overview}%
\label{sec:assumptions-overview}

Our setting has three participants that make the following trust assumptions:
The \emph{service provider} runs a service for clients.  As part of its
operations, the service provider wants to process sensitive client information.

The \emph{client} is a user of the service provider.  It does not trust the
service provider with its sensitive information and demands verifiable
guarantees that the service provider will never see the client's sensitive
information in plain text.

The \emph{enclave provider} makes available enclaves to the service provider.
Both the client and the service provider trust that the enclave provider's
enclaves have the advertised security attributes of integrity, confidentiality,
and verifiability.

We begin with an informal overview of \tool{} to provide intuition.  Subsequent
sections are going to elaborate on the details.  The life cycle of an enclave
application that uses \tool{} involves six steps:

\ding{202} The service provider wants to run an existing service in an enclave.
Assuming that the service builds reproducibly, the service provider then
publishes the service's source code for its clients to audit.

\ding{203} The service provider bundles its service with \tool{} and launches
the enclave, which is now ready to receive incoming connections.

\ding{204} Users audit the service's (freely available) source code.  Once a
user is convinced that the code is free of security bugs, she compiles the
service using the deterministic build system, which results in an image
checksum.

\ding{205} The client establishes an end-to-end encrypted network connection
with the enclave.  Right \emph{after} establishing the secure channel but
\emph{before} revealing any sensitive information, the client provides a nonce
and asks the enclave for an attestation document.

\ding{206} The enclave receives the nonce and asks its hypervisor to generate an
attestation document that contains the client-provided nonce \emph{and} the
public key that the enclave uses to establish the secure channel.  The enclave
returns the resulting attestation document (which contains the image checksum)
to the client.

\ding{207} The client performs various checks (see \S~\ref{sec:attestation} for
details) and trusts the enclave if all checks pass.  The client is then
convinced that it's communicating with the code that the user audited in the
previous steps and is willing to reveal her sensitive information to the
enclave.

\subsection{Enabling reproducible builds}%
\label{sec:build-system}

Once a user audited the service's code, she compiles the code to obtain the PCR0
value (cf.~Table~\ref{tab:pcr}).\footnote{We use the terms ``PCR0 value'' and
``image ID'' interchangeably.}  Crucially, we need a \emph{deterministic
mapping} between the code and its corresponding image ID because the service
provider and clients must agree on the image ID that's running in the enclave.
Docker by itself does not provide a deterministic mapping because---among other
things---Docker records timestamps in its build process, causing subsequent
builds of identical code to result in different image IDs.\footnote{In essence,
a Docker image is an archive of a file system.  A Docker image is reproducible
when separate build processes arrive at the exact same file system, including
meta data like timestamps.}  To obtain reproducible builds, we take advantage of
kaniko~\cite{kaniko}, which is straightforward to integrate into existing
Docker-based workflows.  Kaniko's purpose is to build container images from a
Dockerfile while itself in a container, but we use kaniko because it can do so
reproducibly.  As long as the client and service provider use the same source
code, kaniko version, and compiler, they can build identical images---even when
compiling the code on different platforms, like macOS and Linux.  Equipped with
a locally-generated PCR0 value (henceforth simply called ``image ID''), the
client is now ready to interact with the enclave.

\subsection{\Tool{}'s components}%
\label{sec:framework}

Having discussed how the client and service provider can independently arrive at
identical image IDs, we now turn to \tool{}'s architecture.  The
following sections discuss how \tool{}
communicates securely with the outside world (\S~\ref{sec:networking});
how we facilitate remote attestation (\S~\ref{sec:attestation});
how enclaves can share their key material to allow for horizontal scaling (\S~\ref{sec:sync});
how to thwart side-channel attacks (\S~\ref{sec:side-channels}); and
how to ingest secrets (\S~\ref{sec:secrets}).
Appendix~\ref{sec:example} provides an example of a simple enclave application.

\subsubsection{Enabling seamless and secure networking}%
\label{sec:networking}

\begin{figure}[t]
  \centering
  \begin{tikzpicture}[node distance=20pt] 
  \node [draw,
         label={[anchor=north]above:EC2 host},
         minimum height=100pt,
         align=center,
         minimum width=60pt] (ec2) {};

  \node [draw,
         label={[anchor=north]above:Enclave},
         right=0pt of ec2,
         fill=black!10,
         minimum height=100pt,
         minimum width=100pt] (enclave) {};

  \node [draw,
         below=0pt of enclave.south west,
         xshift=20pt,
         minimum width=160pt] (hypervisor) {Hypervisor};

  \node [draw,
         align=center,
         left=of ec2.east] (proxy) {Proxy};

  \node[draw,
        align=center,
        fill=white,
        xshift=-10pt,
        right=of ec2.east] (tap) {tap};

  \node[draw,
        align=center,
        fill=white,
        yshift=10pt,
        right=of tap.north east] (nitriding) {Nitriding};

  \node[draw,
        align=center,
        fill=white,
        below=of nitriding.south] (app) {App.};

  \node [draw,
         left=of ec2.west,
         xshift=10pt] (letsencrypt) {Let's Encrypt};

  \node [draw,
         above=of letsencrypt] (backend) {Back end};

  \node [draw,
         below=of letsencrypt] (client) {Client};

  \node [right,
         align=center,
         yshift=10pt,
         rotate=90] at (enclave.west)
        {\footnotesize \color{gray} VSOCK\\\footnotesize \color{gray} interface};

  \draw[-latex, densely dashdotted] (app.south) -- ([xshift=51pt]hypervisor.north)
      node [midway, fill=white, circle, inner sep=-2pt] {\ding{202}};

  \draw[latex-latex, densely dotted] ([yshift=5pt]nitriding.west) -- ([yshift=5pt]tap.east)
        node [midway, fill=white, circle, inner sep=-2pt] {\ding{203}};
  \draw[latex-latex, densely dotted] (tap.west) -- (proxy.east);
  \draw[latex-latex, densely dotted] (proxy.west) -- (letsencrypt.east);

  \draw[-latex] (client.east) -- ([yshift=-5pt]proxy.west)
        node [midway, fill=white, circle, inner sep=-2pt] {\ding{204}};
  \draw[-latex] ([yshift=-5pt]proxy.east) -- ([yshift=-5pt]tap.west);
  \draw[-latex] (tap.east) -- ([yshift=-5pt]nitriding.west);
  \draw[-latex] (nitriding.south) -- (app.north);

  \draw[-latex, densely dashed] (app.west) -- ([yshift=-5pt]tap.east)
        node [midway, fill=white, circle, inner sep=-2pt] {\ding{205}};
  \draw[-latex, densely dashed] ([yshift=5pt]tap.west) -- ([yshift=5pt]proxy.east);
  \draw[-latex, densely dashed] ([yshift=5pt]proxy.west) -- (backend.east);
\end{tikzpicture}
  \caption{An architectural diagram illustrating the data flow as \tool{} first
    boots and as clients talk to the enclave application.}%
  \label{fig:networking}
\end{figure}
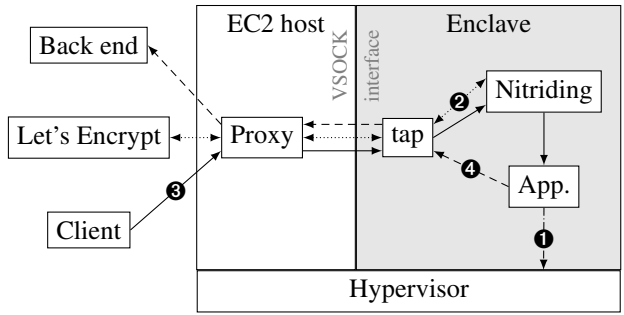

Nitro Enclaves have no networking interface.  Their only way to talk to the
outside world is a VSOCK interface connected to the EC2 host.  \Tool{}
works around this limitation by creating a TAP interface~\cite{tun-tap} inside
the enclave as illustrated in Figure~\ref{fig:networking}.  The TAP interface is
a virtual networking interface that acts as a network bridge between the enclave
and the EC2 host.  This interface routes traffic to a cooperating proxy
application running on the EC2 host, which provides the enclave application with
seamless Internet connectivity for both IPv4 and IPv6.  The proxy supports port
forwarding to both \tool{} and the enclave application (which are independent
processes), allowing clients to talk to either directly.  \Tool{} supports two
modes for the enclave application to receive connections from clients:

{\bf Reverse proxy}: \Tool{} acts as an HTTP reverse proxy and forwards incoming
HTTP requests to the enclave application.  \Tool{} terminates the TLS
connection, meaning that the enclave application can use plain HTTP.  This mode
is only applicable if the enclave application exposes an HTTP API.

{\bf Direct}: In this mode, the enclave application receives incoming
connections directly from the cooperating proxy.  \Tool{} is not involved.  This
is useful for enclave applications that speak protocols other than HTTP, or
require greater flexibility than what the reverse proxy provides.

Having established how the enclave application can send and receive network
packets, we now turn our attention to secure channels; specifically: how can
clients establish a secure channel that is terminated \emph{inside the enclave}?
Enclave applications that receive connections in ``direct'' mode must implement
their own secure channel.   For enclave applications that take advantage of the
``reverse proxy'' mode, \tool{} offers a secure channel in the form of HTTPS.
When the enclave first starts, \tool{} fetches a CA-signed certificate from
Let's Encrypt using the ACME protocol~\cite{acme-protocol} and its TLS-ALPN-01
challenge~\cite{tls-alpn} (step~\ding{203}).\footnote{Unlike the DNS-01 and the
HTTP-01 challenge, TLS-ALPN-01 works entirely in the context of TLS and does not
rely on other ports or protocols, which simplifies deployment.}  Crucially, this
certificate \emph{lives and dies} inside the enclave and its private key cannot
be extracted (or injected) by the service provider because enclaves are sealed
at runtime.
The EC2 host (which is untrusted as per our threat model) can obtain a
CA-signed certificate for the same FQDN because the enclave and the EC2 host
share an IP address.  This however is of little use to the EC2 host because we
tie the enclave's certificate to an attestation document as we will discuss
next.

\subsubsection{Authenticating secure channels}%
\label{sec:attestation}

Assume a client established a TLS connection with an enclave.  How does the
client know that the TLS session is terminated inside the enclave and not by
the EC2 host?  Our trust assumptions state that all parties trust the enclave
provider, Amazon.  We use an enclave's attestation document as the root
of trust and therefore authenticate a secure channel by \emph{binding it to the
enclave's attestation document}.  By including the enclave application's public
key in the attestation document, clients know that they are talking to \emph{an
enclave}.  And by auditing the enclave code and building it reproducibly,
clients know that they are talking to \emph{their trusted enclave}.

We now discuss how we allow clients to retrieve and verify an enclave's
attestation over the Internet because Nitro Enclaves only allow for local
attestation.  After the client established a secure channel with the enclave, it
needs to know that (i) the channel it just established is terminated inside the
enclave (instead of by the EC2 host) and (ii) the enclave is running the code
that the user audited in the previous step.  To that end, the client requests
the enclave's attestation document---a hypervisor-signed document that attests
to the image ID that the enclave is running.  The client begins by provides a
\emph{nonce}---a 160-bit random value---whose purpose is to prevent the service
provider from replaying outdated attestation documents.  Phrased differently,
the client provides a nonce to convince itself that it's talking to a live
enclave.  \Tool{} exposes an HTTP endpoint that clients use to request an
attestation document.  An example request looks as follows:

\begin{lstlisting}[numbers=none,basicstyle=\small\ttfamily]
curl -i "https://enclave/attestation?nonce=abcd..."
HTTP/2 200
content-type: text/plain; charset=utf-8
date: Fri, 27 Jan 2023 16:58:55 GMT

hEShATgioFkQ9alpbW9kdWxlX2lkeCdpLTA2Y...
\end{lstlisting}

Upon receiving this client request, the enclave requests an attestation document
from its hypervisor via an \texttt{ioctl} system call, which makes use of
/dev/nsm, a device that is available inside the Nitro Enclaves.  As
illustrated in Figure~\ref{fig:attestation}, \tool{} asks the hypervisor
to include both the nonce \emph{and} the public key of the enclave's secure
channel in the attestation document and sends the resulting attestation document
to the client.\footnote{If the enclave application runs in ``reverse proxy''
mode, the public key is a hash over the X.509 certificate; otherwise, it's the
public key of whatever secure channel the enclave application uses.}  Upon
receiving the attestation document, the client then verifies the following in
order:

\begin{figure}[t]
  \centering
  \begin{tikzpicture}[node distance=20pt]

  \draw [dotted] (0,0) -- (0,1.75);
  \draw [dotted] (3,0) -- (3,1.75);
  \draw [dotted] (6,0) -- (6,1.75);

  \node at (0,2) {Client};
  \node at (3,2) {Enclave};
  \node at (6,2) {Hypervisor};

  \draw [-latex]
        (0.1, 1.5) -- (2.9, 1.25) node [midway, fill=white, text centered]
        { $n$ };

  \draw [-latex]
        (3.1, 1.25) -- (5.9, 1) node [midway, fill=white, text centered]
        { $n$, $K_{pub}$ };

  \draw [-latex]
        (5.9, 0.75) -- (3.1, 0.5) node [midway, fill=white, text centered]
        { $A(n, K_{pub})$ };

  \draw [-latex]
        (2.9, 0.5) -- (0.1, 0.25) node [midway, fill=white, text centered]
        { $A(n, K_{pub})$ };

\end{tikzpicture}
  \caption{Clients provide a nonce $n$ when requesting an attestation document
  from the enclave.  The enclave asks its hypervisor for the attestation
  document $A$, providing the client's nonce and its public key $K_{pub}$.  The
  hypervisor responds with the attestation document $A(n, K_{pub})$, which the
  enclave forwards to the client.}%
  \label{fig:attestation}
\end{figure}
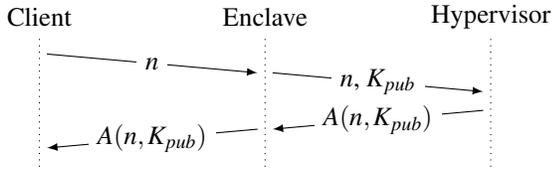

{\bf First}, the attestation document is signed by the AWS root CA whose public
key (which serves as the root of trust) is known to all parties.  This prevents
all parties except Amazon from issuing malicious attestation documents.

{\bf Second}, the challenge nonce is part of the attestation document.  This
prevents adversaries from replaying old attestation documents.

{\bf Third}, the fingerprint of the enclave's X.509 certificate from the TLS
session is part of the attestation document.  This prevents adversaries from
intercepting the secure channel.

{\bf Fourth}, the enclave's image ID is identical to the image ID that the
client compiled locally.  This prevents adversaries from tricking clients into
talking to a malicious enclave.

Only if all four conditions hold is the client convinced that it is talking to
an enclave running the previously-audited code \emph{and} that the secure
channel is terminated inside the enclave.  Note that the EC2 host is able to
intercept the secure channel with its own CA-signed certificate but clients will
only trust the EC2 host if (and only if) it can present an attestation document
that is valid for the enclave image, which it can't because it is unable to
spoof the AWS root CA signature that authenticates the attestation document.
The only way for the EC2 host to obtain such an attestation document is to spawn
an enclave that runs the exact code that the client is expecting---and it
already does exactly that.  Now that the client has established a trust
relationship with the enclave, it is ready to reveal sensitive information to
the enclave.

\subsubsection{Syncing key material among enclaves}%
\label{sec:sync}

Enclaves are sealed at runtime, preventing anyone (including both Amazon and the
service provider) from extracting key material that was generated inside the
enclave.  While a desirable property, this complicates horizontal scaling.  If a
single enclave proves unable to handle traffic load, one must scale horizontally
by starting new enclaves.  In some applications, it is unacceptable for each
enclave to use distinct key material.  Instead, enclaves must synchronize their
key material to appear to the outside world like a single machine.  While it is
possible to accomplish key synchronization using tools like the AWS key
management service (KMS),\footnote{One could encrypt the keys using a KMS policy
that dictates that only enclaves are allowed to decrypt it, and store the
encrypted key in a location that all enclaves can access, e.g., an S3 bucket.}
we refrain from using KMS because users currently cannot verify that a
KMS ``key policy'' is truly immutable.  We therefore devise a novel,
user-verifiable protocol that enables key synchronization without having to rely
on external services.

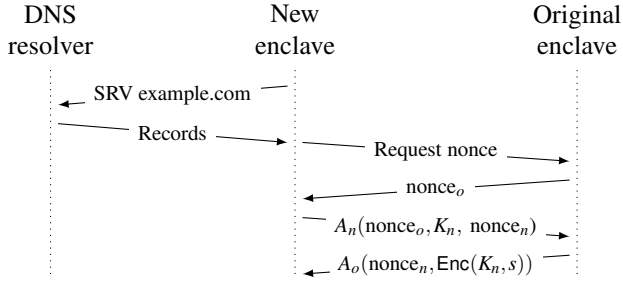
\begin{figure}[t]
  \centering
  \begin{tikzpicture}[node distance=20pt]

  \def\dnsright{0.1}
  \def\newleft{3.15}
  \def\newright{3.35}
  \def\origleft{6.9}

  \draw [dotted] (0,    0) -- (0,    2.75);
  \draw [dotted] (3.25, 0) -- (3.25, 2.75);
  \draw [dotted] (7,    0) -- (7,    2.75);

  \node [align=center] at (0,    3.25) {DNS\\resolver};
  \node [align=center] at (3.25, 3.25) {New\\enclave};
  \node [align=center] at (7,    3.25) {Original\\enclave};

  \draw [-latex] (\newleft, 2.5) -- (\dnsright, 2.25)
        node [midway, align=center, fill=white]
        {\footnotesize SRV example.com};
  \draw [-latex] (\dnsright, 2) -- (\newleft, 1.75)
        node [midway, align=center, fill=white]
        {\footnotesize Records};

  \draw [-latex] (\newright, 1.75) -- (\origleft, 1.5)
        node [midway, fill=white, align=center]
        {\footnotesize Request nonce};
  \draw [-latex] (\origleft, 1.25) -- (\newright, 1)
        node [midway, fill=white, align=center]
        {\footnotesize $\textrm{nonce}_o$};

  \draw [-latex] (\newright, 0.75) -- (\origleft, 0.5)
        node [midway, fill=white, align=center]
        {\footnotesize $A_{n}(\textrm{nonce}_o, K_n, \ \textrm{nonce}_n$)};

  \draw [-latex] (\origleft, 0.25) -- (\newright, 0)
        node [midway, fill=white, align=center]
        {\footnotesize $A_{o}(\textrm{nonce}_n, \textsf{Enc}(K_n, s))$};

\end{tikzpicture}
  \caption{When a new enclave bootstraps, it discovers existing enclaves by
    obtaining the DNS SRV record for its own, hard-coded FQDN.  The enclave then
    initiates key synchronization by first requesting a nonce.  Then, the new
    enclave requests the origin enclave's key material by submitting its own
    attestation document, followed by receiving the origin enclave's attestation
    document, which contains encrypted key material.}
  \label{fig:key-synchronization}
\end{figure}

We solve this problem in two steps: \emph{discovery}, followed by
\emph{synchronization}.  First, enclaves must be able to discover each other,
i.e., learn each other's IP addresses.  Then, enclaves can establish connections
with each other and initiate key synchronization.  Our protocol dictates that
when a new enclave bootstraps, it first tries to discover already-existing
enclaves.  If there are none, the enclave knows that it is the ``origin''
enclave.  It then generates new key material that it will share with future
enclaves.  If however it discovers other enclaves, the new enclave establishes a
connection with another, randomly-chosen enclave and initiates key
synchronization.  Crucially, key material is only shared after \emph{mutual
attestation}, i.e., the origin and new enclaves verify each other, and exchange
key material only if remote attestation succeeds.  Key synchronization happens
in three steps, as illustrated in Figure~\ref{fig:key-synchronization}.

\textbf{First}, once a new enclave is spun up, it queries the DNS SRV record of
the FQDN that is hard-coded in the enclave, e.g., example.com.  The DNS resolver
will return the record, containing a list of enclaves that are already running
and initialized.  The new enclave picks a random enclave from the list and
initiates key synchronization.  Running Nitro Enclaves as part of Kubernetes
can handle DNS record generation automatically.

\textbf{Second}, the new enclave asks the existing enclave for a random nonce,
$\textrm{nonce}_o$.  Both enclaves cache $\textrm{nonce}_o$ for one minute to
mitigate denial-of-service attacks.

\textbf{Third}, the new enclave now requests the key material from the existing
enclave.  As part of the request, it provides its attestation document that
contains $\textrm{nonce}_o$ (to prove freshness to the existing enclave);
$\textrm{nonce}_n$ (the existing enclave is expected to add this nonce to its
attestation document); and $K_n$ (a NaCl public key~\cite{nacl} to which the key
material should be encrypted).  Upon receipt of the new enclave's attestation
document, the existing enclave verifies the attestation document's signature and
ensures that the new enclave is running the same code, i.e., the image ID that
uniquely identifies the enclave image is identical.  Once the existing enclave
is convinced that it is dealing with a genuine new enclave, it creates an
attestation document by including $\textrm{nonce}_n$ (to prove freshness to the
new enclave) and $\textsf{Enc}(K_n, s)$---the key material $s$ is encrypted
using the public key that the new enclave provided in the request.  Finally, the
new enclave verifies the attestation document, decrypts the key material, and
uses it to finish bootstrapping.

The security of key synchronization is paramount.  We take advantage of mutual
remote attestation to protect the key material.  As an optional layer of
defense-in-depth, synchronization should be configured to use a private network,
which prevents arbitrary Internet hosts from talking to the synchronization
endpoint.  While not required, we recommend running enclaves as part of
Kubernetes because it provides a private network that is shared by enclaves.

In their 2022 USENIX Security paper, Chen and Zhang present MAGE, a protocol
that allows enclaves to mutually verify each other without relying on a trusted
third party~\cite{Chen2022a}.  We could have built key synchronization on top of
MAGE but found that our setting is considerably simpler because only
\emph{identical} enclaves request each other's key material, eliminating the
need for the more flexible---but also more complex---MAGE protocol.

\subsubsection{Side-channel attacks}%
\label{sec:side-channels}

The enclave's EC2 host cannot see \emph{what} clients send to the enclave but it
can see \emph{how much} clients send and \emph{how long} it takes the enclave to
process data.  The EC2 host can exploit these side channels to learn more about
the client's confidential information and computation.  While such side channels
must be avoided, \tool{} is not the place to do so.  Instead, it is the enclave
application developer's responsibility to identify and address this class of
attacks, e.g., by implement constant-time processing.

\subsubsection{Ingesting secrets}%
\label{sec:secrets}

A key design requirement of \tool{} is that users must be able to audit the
enclave application's code.  The service provider is therefore unable to hide
any software configuration (e.g., confidential API keys) from the user.  Service
providers can work around this shortcoming by exposing an authenticated HTTP
handler that takes as input arbitrary data that updates the enclave's state.
Consider a system that takes as input client IP addresses, anonymizes them, and
forwards the anonymized addresses to a back end.  The service provider now wants
to compare submitted IP addresses to a confidential deny list.  If however the
deny list is hard-coded in the publicly available enclave application, it is
readily visible to anyone.  The service provider can solve this problem by
adding to the enclave application a new HTTP handler that takes as input the
confidential data it seeks to protect from the users' eyes.  Once the enclave is
running, the service provider loads the confidential deny list.  To prevent
users from submitting bogus data, the endpoint must be authenticated.  One
could accomplish this by hard-coding the service provider's public key in the
enclave application, and only accepting deny lists that carry a valid signature.
Another possibility is to expose this endpoint only to the EC2 host, so only the
service provider has access to it.  We provide an example of this in
Section~\ref{sec:vct}.

This technique for ingesting secrets is flexible---so flexible, in fact, that
the service provider could abuse it to ingest code at runtime, which would
nullify the enclave's verifiability requirement.  Vigilant users would never
trust an enclave whose code can change at runtime.  We therefore argue that an
HTTP handler for the purpose of ingesting secrets must be constrained so that
only well-defined data of a certain type (like a deny list) can be ingested.

\section{\Tool{}-based applications}%
\label{sec:applications}

In this section, we build three applications on top of \tool{}, each a research
contribution in its own right.
First, we build an application that allows a service provider to disclose its
infrastructure configuration in a user-verifiable way, thus eliminating the
trust that users have to have in third-party infrastructure (\S~\ref{sec:vct}).
Second, we launch a Tor bridge inside an enclave, which mitigates several
classes of attacks that the Tor network has struggled with in the past
(\S~\ref{sec:tor-bridge}).
Third, we show that \tool{} can handle computationally expensive workloads by
moving a Web browser into an enclave and letting users interact with it via a
remote desktop environment (\S~\ref{sec:browser}).

\subsection{Verifiable configuration transparency}%
\label{sec:vct}

Service providers typically outsource their infrastructure to third-party
providers like content delivery networks or cloud computing vendors.  The way
these third-party providers are configured often affects user privacy.  For
example, a service provider may configure a third-party reverse proxy to strip
client IP addresses before requests are forwarded to the servers that are under
the service provider's control.  How can users know that the reverse proxy is in
fact configured to strip client IP addresses?  We built an enclave application
that solves this problem by disclosing infrastructure configuration in a
user-verifiable way.  The idea, illustrated in Figure~\ref{fig:vct} in
Appendix~\ref{sec:more-diagrams}, consists of a lightweight enclave application
whose sole purpose is to answer client requests by querying the API of the
third-party infrastructure provider.  We built our proof-of-concept
implementation for Cloudflare but the code is easily adapted for other
providers.

To interact with Cloudflare's API endpoint, one needs a confidential bearer
token for authentication and a semi-confidential zone ID~\cite{spectrum-config}.
Unlike the API endpoint's URL, these two values cannot be hard-coded in the
(public) source code of the enclave application.  We therefore add a second Web
server to the enclave application whose only purpose is to receive as input the
bearer token and the zone ID (cf.  \S~\ref{sec:secrets}).  We carefully
constrained this Web server's HTTP handlers, making it impossible to inject
anything into the enclave \emph{but} the bearer token and the zone ID.  We
further configured the proxy to only forward to this Web server connections
originating at the EC2 host.  Internet-connected adversaries cannot reach this
Web server.  After the administrator launches the enclave, she injects the
confidential values into the enclave by calling the private HTTP endpoint from
the EC2 host.  Users have no reason to be concerned about this secret endpoint
because the enclave application's source code shows that the secret values are
only used for the API request to Cloudflare.  Clients talk to this enclave
application via a single endpoint, which returns the JSON-encoded Cloudflare
configuration in the HTTP body, and the enclave's attestation document in a
custom HTTP header.  Section~\ref{sec:attestation} explained that the
attestation document must contain a client-provided nonce.  The client provides
this nonce as part of the HTTP request URL.  In summary, clients make the
following request:

\begin{lstlisting}[numbers=none,basicstyle=\small\ttfamily]
GET /verify?nonce=3a26d...a937f HTTP/2
Host: enclave.example.com
\end{lstlisting}

And the server responds with:

\begin{lstlisting}[numbers=none,basicstyle=\small\ttfamily]
HTTP/2 200 OK
Date: Mon, 13 Mar 2023 18:34:29 GMT
Content-Type: application/json
X-Attestation-Document: hEShATgioFkRJalpbW9kdWx...

{
  "domain": "service.provider.com",
  "modified_on": "2021-09-08T18:04:10.711156Z",
  ...
}
\end{lstlisting}

Upon receiving the enclave's response, the client first verifies the
authenticity of the attestation document (cf. \S~\ref{sec:attestation}).  Once
convinced that the enclave's response is authentic, the client inspects the
body of the response---which comes directly from Cloudflare's API.  In
particular, the client consults Cloudflare's API documentation to verify that
the service provider configured Cloudflare as promised.  Finally, the user
verifies that the domain inside the enclave's response matches the domain that
the service provider makes available to its users.

\subsection{Tamper-resistant Tor bridge}
\label{sec:tor-bridge}

The Tor network's security rests on the assumption that certain relays in a
user's circuit do not collude.  This assumption does not always hold, as in the
2014 attack that sought to deanonymize onion service
users~\cite{Dingledine2015a}.  The attack consisted of several malicious relays
that injected a sequence of Tor cells to encode a messages along the
circuit~\cite[\S~5.6]{tor-spec}.  Being an active attack, this required a
modified relay that deviated from the Tor protocol.  If done well, such attacks
can be difficult to spot.

\Tool{} can help mitigate such attacks by running Tor infrastructure inside an
enclave.  By taking advantage of remote attestation, Tor clients can rest
assured that they are communicating with an authentic Tor implementation that
does not deviate from the Tor protocol.  We demonstrate that this is possible
by setting up a Tor bridge inside an enclave.\footnote{We chose to set up a Tor
bridge instead of a relay because bridges can be configured to remain private
and therefore cause no harm to the network in case our implementation had
bugs.}

\textbf{Proof-of-concept deployment}: Running the Tor executable inside an
enclave is straightforward: the Dockerfile and startup script are nearly
identical to Listing~\ref{fig:example}.  Remote attestation however is more
complicated.  Ideally, Tor clients would attest the authenticity of their Tor
bridge as part of the Tor protocol itself but for the sake of this prototype, we
are content with handling remote attestation outside the Tor protocol: Once the
Tor bridge is done bootstrapping, it registers its long-term identity key with
\tool{}.  The enclave then exposes two TCP ports: port 443 for \tool{} and port
9001 for Tor.  A Tor client first remotely attests the enclave, followed by
establishing a circuit over the bridge and---while establishing the
circuit---verifies that the bridge's long-term identity key is identical to the
key in the attestation document.  If so, the client can rest assured that it's
talking to a publicly verifiable Tor bridge.

We implemented the aforementioned prototype and configured Tor Browser v12.0.1
to use our in-enclave bridge.  Using this setup, we were able to watch 4K
YouTube videos without buffering delays.
The above setup works well for an ad-hoc setup but is insufficient for
network-wide deployment of in-enclave relays and bridges.  In this case, relays
and bridges need a way to announce if they support relay attestation.  This is
the job of the Tor network's consensus, which is generated every hour by the
distributed directory authorities.  Changes to the network consensus are
complex and need to be addressed in the protocol specification and Tor's
reference implementation.

\textbf{Comparison to SGX-Tor}: In their NSDI'17 paper, Kim et al. augmented the
Tor code with Intel SGX, thus giving clients, relays, and directory authorities
the ability to remotely verify each other~\cite{Kim2017a}.  Our approach differs
in the following aspects:
Clients that seek to verify an SGX enclave's attestation document need to talk
to Intel's attestation service, which brings with it an array of privacy
problems~\cite[\S~1.2]{Chen2019a}.  With Nitro Enclaves, clients can verify an
attestation document without having to contact a third party, provided that they
have a copy of Amazon's root CA public key.
Next, the authors envision the entire Tor network to take advantage of SGX,
which is not feasible in our approach: Nitro Enclaves can only run in
Amazon-controlled AWS.  If all Tor relays ran inside AWS, Amazon would see both
traffic entering and exiting the network---ideal conditions for end-to-end
correlation attacks.  We therefore believe that only select Tor bridges benefit
from running in \tool{}, lest anonymity is jeopardized.
As for practicality, Kim et al. had to go to great lengths to patch Tor to
support SGX~\cite[\S~5]{Kim2017a}.  Our proof-of-concept implementation took one
afternoon.  Finally, since Kim et al.'s paper was published, Intel announced the
discontinuation of SGX support for consumer-grade Core CPUs, which further
limits the number of SGX-capable Tor clients.

\subsection{An in-enclave Web browser}%
\label{sec:browser}

The previous enclave application focused on a low-latency, high-throughput use
case.  We now build an application that requires low latency in addition to
being \emph{computationally demanding}: we move a modern Web browser into an
enclave, thus isolating the browser from its user's desktop environment.
If a browser is compromised by, say, a malicious Web site, then the malicious
code is constrained to the enclave, unable to interact with the user's desktop
environment.  In addition, users benefit from not having to provide the
ever-increasing computational resources to run the browser.
The downside of this approach is that the enclave's EC2 host gets to see the
browser's traffic---but this can be solved by using a VPN or Tor.
Alternatively, the pervasive deployment of HTTPS helps protect page contents
from the EC2 host's prying eyes and one could configure the browser to use
DNS-over-HTTPS to further protect DNS traffic.

We start with an Ubuntu Docker image, which we extend by installing basic X11
graphical utilities, OpenSSH, TigerVNC, the i3 window manager, and a Chromium
browser, all via Ubuntu's package manager.\footnote{We used Ubuntu 20.04.4 from
docker.io.}  We used OpenSSH port forwarding to tunnel TigerVNC's TCP traffic,
which prevents the EC2 host from spying on VNC traffic.  The enclave's SSH
public key acts as the root of trust, and it lives and dies inside the enclave.
We configure \tool{} to add a hash over this public key to the attestation
document.  Before establishing an SSH connection, clients fetch the enclave's
attestation document from the \tool{} port to learn what SSH public key to
expect.
By default, AWS's tooling for Nitro Enclaves assumes applications smaller than
what we need for a remote desktop environment.  Depending on the instance size
and tool version, we sometimes had out-of-memory errors converting container
images to enclave images.  Additionally, the enclave manager allows a maximum
allocation of 512 MB to each container, which isn't enough to run most graphical
desktop software by default.  This limit needs to be raised.

There is plenty of room for improvement, both in reducing the enclave's image
size and in reducing the latency that our VNC server induces.  One can also take
advantage of AWS's numerous data centers by launching the browser enclave in a
region that is close to the user, to further minimize latency.

\textbf{Subjective user experience}:
Upon using a VNC client to interact with the in-enclave Web browser, we found
that navigation was relatively painless but large screen updates were spread
over a noticeable amount of time, usually obscuring the underlying animation.
Video played at less than full framerate, and VNC doesn't support audio
playback.  However, the user-perceptible latency is dominated by the round-trip
time between the client and the enclave.  In any case, this application
demonstrates that the enclave itself is no barrier to achieving similar
performance to any other remote desktop service.

\textbf{Alternative approaches}:
Wang et al.'s WebEnclave work sets out to protect Web sites from malicious
browser extensions: Web developers can use the \texttt{<web-enclave>} tag to
instruct the browser to execute code inside WebEnclave~\cite{Wang2021a}.  Unlike
WebEnclave, our application protects the user's operating system from the
browser, instead of Web pages from browser extensions.

\section{Evaluation}%
\label{sec:evaluation}

We now evaluate \tool{} with respect to security, financial cost, and
performance.  We ran all our performance measurements on a c5.xlarge EC2
instance~\cite{c5-instance}, which is on the lower end of enclave-capable
instance types.  Our performance numbers therefore represent a lower bound of
what's possible with Nitro Enclaves.  More powerful instance types will yield
better performance.

\subsection{Security considerations}%
\label{sec:security}

There are three key components to the overall security of a \tool{} application:
(i) Amazon's Nitro system itself,
(ii) \tool{}, and
(iii) the application running on top of \tool{}.

The security foundation lies in the soundness of the design of Nitro Enclaves.
While Amazon published the conceptual design~\cite{Bean2022a}, the concrete
hardware and software implementation is closed source.  The decision to
allocate physically separate resources to enclaves is sound but only time will
tell if Nitro Enclaves can resist the types of attacks that have been plaguing
SGX.

\Tool{}'s security reduces to the complexity of our code and the security of its
cryptographic building blocks.  These building blocks are SHA-256 (to hash
public key material that is embedded in the attestation document), the NaCl
cryptographic library~\cite{nacl} (to implement key
synchronization),\footnote{Specifically, we use NaCl's box API, which uses
Curve25519, XSalsa20, and Poly1305 to encrypt and authenticate messages.} TLS in
at least version 1.2 (to provide a secure channel between clients and the
enclave), and Go's CSPRNG, which is seeded with randomness from the Nitro
hypervisor.
One measure of our code complexity is its size.  Excluding unit tests, \tool{}
counts less than 1,700 lines of code and has nine direct dependencies
that are not maintained by either us or the Go project.\footnote{Most
dependencies provide networking functionality like a user-space TCP stack, code
that provides a TAP interface, and a wrapper for Linux's netlink interface.}
Nine is worse than zero, but is still manageable and auditable in its entirety.
Our choice of using the memory-safe Go and the (comparatively) small trusted
computing base reduces---but does not eliminate!---the attack surface.

The highest layer in the software stack is the enclave application itself.  The
most significant security issues are side channel attacks and programming bugs.
It is the application developer's responsibility to prevent side channel attacks
and write bug-free code.  As we pointed out in Section~\ref{sec:limitations},
programming bugs can be intentional, i.e., the service provider may deliberately
introduce bugs that leak sensitive information.  From the user's point of view,
eternal vigilance is therefore the price of security.

\subsection{Financial cost}%
\label{sec:cost}

Nitro Enclaves do not incur any extra cost in addition to that of the underlying
EC2 host---they can be considered a ``free'' extension to EC2.  Nitro enclaves
are however only available for select types of EC2 instances because they
require their own CPU and a minimum amount of memory, and those instance types
are pricier than the lowest tier that AWS offers.  We tested all of the
practical applications of Section~\ref{sec:applications} on a c5.xlarge
instance, which is on the lower end of enclave-enabled EC2 instance types.  This
instance comes with four vCPUs and 8 GiB of memory.  As of March 2023, a
c5.xlarge instance costs USD 0.17 per hour, which amounts to approximately USD
125 per month.

\subsection{Attestation document request rate}%
\label{sec:attestation-performance}

\begin{figure}[t]
    \centering
    \input{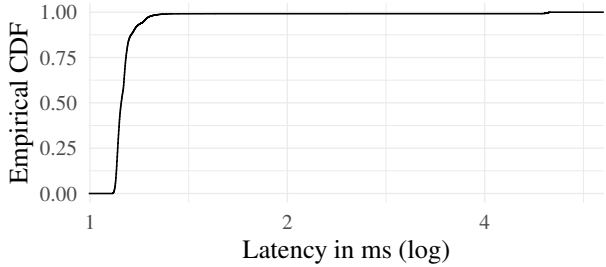}
    \caption{The latency distribution (as CDF) of requesting 51,821
    attestation documents from the Nitro hypervisor.}\label{fig:att-perf}
\end{figure}

The fetching of attestation documents is a critical part of our framework's
overall performance.  We built a stress test enclave application that runs a
busy loop for 60 seconds to request as many attestation documents as possible.
For each request, we ask the hypervisor to include an incrementing nonce in the
attestation document to eliminate any speedups by caching.
Figure~\ref{fig:att-perf} illustrates the results.  We were able to obtain
approximately 860 documents per second with the median request taking 1.1 ms to
complete.  99\% of requests finished in less than 1.31 ms.  Our experience from
building enclave applications (cf.~\S~\ref{sec:applications}) suggests that the
attestation document request rate is unlikely to be a bottleneck for real-world
deployments.

\subsection{Application latency}%
\label{sec:latency}

We seek to measure two types of latency: (i) the latency induced by the
interface between an EC2 host and its Nitro Enclave and (ii) the latency induced
by \tool{}, both with and without \tool{}'s reverse proxy configuration.  The
former is outside our control while the latter is entirely controlled by us.  We
built a lightweight enclave application to measure these latencies.  The
application implements a Web server written in Go that responds with the string
``hello world'' upon receiving requests for its index page.  We made this
application minimal because we're only interested in the latency \emph{before} a
request reaches the enclave application.  For this reason, the Web server only
speaks the computationally inexpensive HTTP (instead of HTTPS).  To simulate
clients, we use the HTTP load test tool Baton~\cite{baton} in git commit
\texttt{576339}.  We patched Baton's source code to add VSOCK support (to send
requests directly to the enclave, via the VSOCK interface) and to log latency
percentiles.  Equipped with both an HTTP server and a client, we measure HTTP
request latencies for five setups:

\begin{description}
  \item[Loopback] The client talks to the Web server via the loopback interface.
    No enclave is involved.  This setup constitutes the latency baseline we
    compare against.

  \item[Enclave] The Web server runs inside a Nitro Enclave but \emph{without
    \tool{}}.  All traffic goes over the VSOCK interface.  This measures the
    latency that the Nitro Enclave's VSOCK interface introduces.

  \item[\Tool{}-nrp] The Web server runs inside a Nitro Enclave but without a
    reverse proxy.  The suffix ``nrp'' is short for ``no reverse proxy''.  This
    measures the latency introduced by \tool{}'s TAP forwarding code.

  \item[\Tool{}] The Web server runs inside a Nitro Enclave with \tool{} acting
    as a reverse HTTP proxy.  This measures the latency introduced by \tool{}'s
    TAP forwarding code \emph{and} its reverse HTTP proxy.
\end{description}

We run Baton on the parent EC2 host and instruct it to send 100,000 requests in
six sequential experiments, using 1, 5, 10, 25, 50, and 100 concurrent threads.
Note that our measurements are designed to measure the \emph{lower bound} for
latency.  Real-world applications will exhibit higher latency because clients
send their requests over the Internet (which adds considerable networking
latency) and the enclave application is likely to be more complex (which adds
computational latency).

\begin{figure}[t]
  \centering
  \input{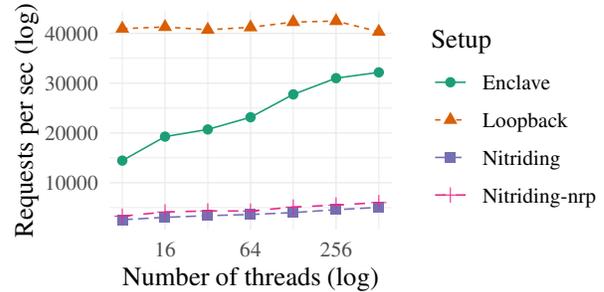}
  \caption{The number of HTTP requests per second as requests are sent from an
  increasing number of threads.}%
  \label{fig:http-reqs-sec}
\end{figure}

Figure~\ref{fig:http-reqs-sec} illustrates the results.  In all setups, the
requests per second increase as the number of threads increases, but only up to
25 threads, at which point we see diminishing returns.  As expected, the
loopback interface---our baseline---performs the best, handling 41,000 requests
per second for 25 threads.  The Enclave setup can sustain approximately half of
that, namely 20,000 reqs/s.  Recall that the Enclave setup constitutes the
maximum achievable performance for \tool{}.

We find that both \tool{} and \tool{} without a reverse proxy (denoted as
\Tool{}-nrp) can sustain 2,300 and 2,600 reqs/s, respectively---13\% of what is
achievable over the enclave interface.  As expected, \tool{} performs better
without reverse proxy because less complexity is involved.  We attribute the
performance difference between \tool{} and the Enclave baseline to the
user-space TCP stack that our software dependency gVisor introduces.

\begin{figure*}[t]
  \centering
  \input{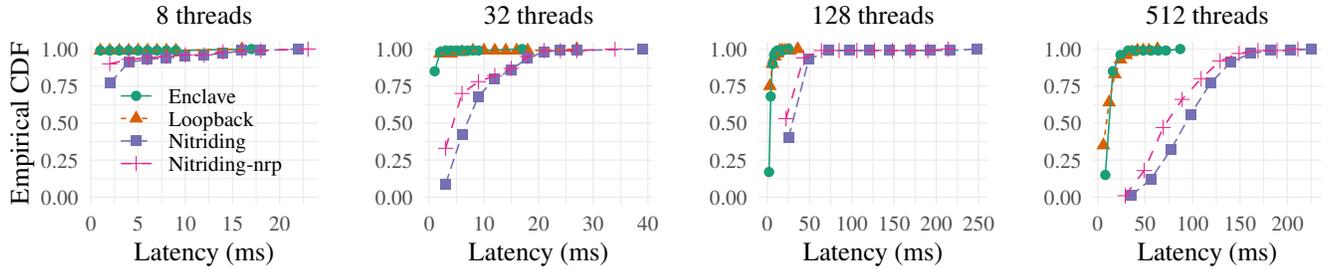}
  \caption{The round-trip time distributions (as CDF) of stress-testing an
  in-enclave Web server as the number of concurrent requesters increases from 1
  to 100 threads.}%
  \label{fig:http-req-lat}
\end{figure*}

Figure~\ref{fig:http-req-lat} shows how long it takes to answer the HTTP
requests we issued as part of this experiment.  The four charts show a latency
CDF for 1, 10, 50, and 100 threads, respectively.  We omitted charts for 5 and
25 threads because of page constraints.

\subsection{Application throughput}%
\label{sec:throughput}

Next, we measure the throughput that we can achieve over the VSOCK interface.
To that end, we use a VSOCK-enabled fork of the iperf3 performance measurement
tool in git commit \texttt{9245f9a}~\cite{iperf-vsock}.  iperf3 measures the
throughput of a networking link using a client/server model.  In our experiment,
we start an iperf3 server instance inside the enclave and the corresponding
client instance on the parent EC2 host.\footnote{The command that we ran on the
server was ``\texttt{iperf3 -{}-vsock -s}'' and on the client ``\texttt{iperf3
-{}-vsock -c 4}.''}  The client then talks to the server via the VSOCK interface
and determines the throughput.  Table~\ref{tab:iperf3} shows the results of this
experiment.  When running both the iperf3 client \emph{and} server on the EC2
host---which effectively measures the throughput of the EC2 host's loopback
interface---we achieve 57 GBit/s of throughput.  Running the iperf3 server
inside an enclave limits throughput to 3 Gbit/s while \tool{} results in
approximately 0.3 Gbit/s.  Iperf3 does not use HTTP and we therefore cannot
measure \tool{} in its reverse proxy mode.

\begin{table}[t]
    \centering
    \begin{tabular}{l r r}
    \toprule
      Setup & C $\rightarrow$ S (Gbits/s) & S $\rightarrow$ C (Gbit/s) \\
    \midrule
      Loopback        & 57.0 & 57.0 \\
      Enclave         &  3.6 &  3.2 \\
      Nitriding-nrp &  0.3 &  1.1 \\
    \bottomrule
    \end{tabular}
    \caption{The TCP throughput of running iperf3 over the
    loopback interface, inside an enclave, and inside an enclave using
    nitriding (without reverse proxy).}%
    \label{tab:iperf3}
\end{table}

\section{Limitations}%
\label{sec:limitations}

An obvious limitation of \tool{} is its reliance on Amazon, which acts as the
root of trust.  Our trust assumptions state that all parties must trust Amazon.
Placing one's trust in a single corporation's proprietary technology is
problematic but this is a common limitation of enclaves---SGX-based applications
must trust Intel while TrustZone-based applications must trust ARM.

Next, our system relies on at least some users auditing the service provider's
enclave application.  Needless to say, not many users have the skills to audit
code.  In fact, even among programmers, only a fraction may be qualified to
audit source code for vulnerabilities.  So what are the non-programmers to do?
We envision users to congregate in forums where matters related to the service
provider are discussed.  A tech-savvy subset of users is going to organize code
reviews and publish their findings.  Non-technical users may then trust the
users who audited the source code.  This is no different from other free
software projects: nobody audits all the software that they use, ranging from
the kernel to the myriad of user space applications.

The Underhanded C Coding Contest's~\cite{underhanded-c} goal was the
implementation of benign-looking code that was secretly malicious.  The contest
attracted numerous impressive submissions which showed that it is surprisingly
difficult to find bugs \emph{even if one knows} that there is a bug in a given
piece of code.  Analogously, the service provider could try to hide subtle, yet
critical bugs in the code to exfiltrate information from the enclave.  On top of
that, if the service provider ever gets caught, it may have plausible
deniability and pretend that the exfiltration bug was an honest programming
error.  We are unable to solve this class of attacks but we can mitigate it
by keeping the trusted computing base small.

\section{Related work}%
\label{sec:related-work}

Arnautov et al.\ present in their OSDI'16 paper a mechanism that allows Docker
containers to run in an SGX enclave~\cite{Arnautov2016a}---conceptually similar
to Nitro Enclaves, which are effectively compiled Docker images.
In their 2022 arXiv report, King and Wang~\cite{King2022a} propose HTTPA---an
SGX-based extension to HTTP that makes a Web server attestable to clients.  Our
framework also allows for attestable Web services, but without modifications to
HTTP.

\textbf{Applications of enclaves}:
Researchers have proposed numerous and diverse enclave-enabled systems, ranging
from DeFi oracles~\cite{Zhang16a}, to health apps for
COVID-19~\cite{Mailthody21a}, to networking middleboxes~\cite{Han17a}.  Despite
avid interest in academia, large-scale, real-world deployments of enclaves are
sparse.  In 2017, the Signal secure messenger published a blog post on private
contact discovery~\cite{Marlinspike17a}, which makes it possible for Alice to
discover which of the contacts in her address book use Signal without revealing
her contact list.  The Signal team accomplished this by relying on an SGX
enclave that runs the contact discovery code.  Two years later, in 2019, the
Signal team built its ``secure value recovery'' feature on SGX as
well~\cite{Lund19a}.

\textbf{Frameworks for enclave development}:
To facilitate working with enclaves, several frameworks have emerged that
abstract away complicated and error-prone low-level aspects of enclaves.
Examples are Asylo~\cite{asylo} and Open Enclave~\cite{openenclave}---both
libraries are implemented in C/C++ and are hardware agnostic, meaning that the
``enclave backend'' can be switched from, say, TrustZone to SGX.  While
frameworks render enclave development more convenient, memory unsafe languages
like C and C++ make it dangerously easy to introduce memory corruption bugs
that jeopardize the security of the enclave~\cite{Lee2017a}.  Cognizant of
this issue, Wang et al.\ implemented a performant Rust layer on top of Intel's
C++-based SGX SDK, making it possible to develop memory-safe applications in
SGX~\cite{Wang2019a}.

\Tool{} is built in the memory-safe Go programming language, which
eliminates an entire class of bugs that could jeopardize the security of
enclave applications, and unlike Asylo and Open Enclave, \tool{} only
supports Nitro Enclaves because the security guarantees of a framework are only
as strong as the underlying enclave hardware, and in the case of Intel, ARM,
and AMD, side channel attacks remain a serious concern.

\textbf{Nitro Enclaves versus Intel SGX}:
We now make an attempt to compare how Nitro Enclaves and SGX differ in their
threat model, their development model, and in the way they can address security
vulnerabilities.

\emph{Threat model}:
Both Nitro Enclaves and SGX protect against compromise of the host operating
system.  SGX further protects against compromise of any component other than the
CPU itself, which includes---if present---the hypervisor.  Specifically, SGX
assumes that there are no flaws in the CPU's silicon or microcode, and the
private key is not compromised.  While not explicitly stated, Amazon's design
document suggests that Nitro Enclaves assume that the Nitro system (including
the Nitro card, the security chip, and the hypervisor) is trusted.  Both Nitro
enclaves and SGX assume that side channel attacks are not feasible.  For SGX,
this assumption has not held~\cite{Nilsson20a,Fei2021a}.

\emph{Development}:
Intel's SGX was not designed to seamlessly move entire applications into the
context of an enclave because the libc that is provided by Intel's SDK lacks
support for many functions and system calls.  Instead, application developers
were meant to partition their application, i.e., move trusted code fragments
into the enclave while the remaining code ran outside the enclave.  However,
projects like Haven~\cite{Baumann2014a} and SCONE~\cite{Arnautov2016a} made it
possible to run entire unmodified applications inside an SGX enclave.  Nitro
enclaves in contrast provide by default what Arnautov et al.\ developed in their
OSDI'14 paper~\cite{Arnautov2016a}: a way to seamlessly run a Docker container
inside an enclave.

\emph{Addressing vulnerabilities}:
What means do Intel and Amazon have to mitigate attacks against their enclave
technology?  Amazon is in possession and control of all hardware and software.
A hardware flaw in Nitro cards may prove expensive and complicated to fix but a
fix is feasible without involving the customer.  Intel has less flexibility
considering that their CPUs are under customer possession.  Some SGX
vulnerabilities have been addressed by updating CPU microcode, which may be a
standard procedure for cloud providers but certainly less so for end users.

Finally, as of March 2023, Intel is in the process of rolling out their
``Trusted Domain Extensions'' processor feature, which is conceptually similar
to Nitro Enclaves in the sense that it aims to protect virtual machines from
both the hypervisor and all other software, including the operating
system~\cite{tdx}.

\textbf{Attacks against enclaves}:
Enclaves based on Intel's SGX technology share a CPU with untrusted code, which
raises the flood gates for side channel attacks.  Consequently, attacks have
taken advantage of
speculative execution~\cite{VanBulck2018a,VanSchaik2021a},
branch ``shadowing''~\cite{Lee2017b},
the interface between SGX and non-SGX code~\cite{Bulck19a},
software faults~\cite{Murdock2020a},
shared caches~\cite{Brasser2017a},
and memory management~\cite{Wang2017a}.
Despite the considerable number of practical attacks, there is opportunity to
strengthen SGX against side channel attacks.  Oleksenko et al.\ introduce in
their ATC'18 paper a system that protects unmodified SGX applications from side
channel attacks by executing the enclave code on a CPU separate from the
untrusted code.  Note that this is the default for Nitro Enclaves.

For a comprehensive overview of attacks against SGX, refer to Fei et al.'s
survey~\cite{Fei2021a}, Nilsson et al.'s arXiv report~\cite{Nilsson20a}, and
Van Schaik et al.'s technical report~\cite{Schaik2022a}.

Among all currently-available commodity enclaves, Intel's SGX has received the
most attention from academia but ARM's TrustZone and AMD's SEV have not been
spared and share SGX's conceptual security flaws.  In a CCS'19 paper, Ryan
demonstrates an attack that exfiltrates ECDSA private keys from Qualcomm's
implementation of a hardware-backed keystore which is based on
TrustZone~\cite{Ryan2019a}.  Similarly, Li et al.\ showed in a USENIX
Security'21 paper how an attacker can exfiltrate private keys from AMD
SEV-protected memory regions.  In a CCS'21 paper, Li et al.\ showed how an
attacker-controlled VM can read encrypted page tables, and how an attacker can
create an oracle for encryption and decryption.

While Nitro Enclaves are still young and have received nowhere near the same
scrutiny as SGX and friends, we believe that their dedicated hardware resources
provides stronger protection from side channel attacks than enclaves
that are based on shared CPU resources.

\section{Conclusion}%
\label{sec:conclusion}

This work presents \tool{}, a tool kit that facilitates the rapid development of
flexible, powerful, and secure enclaves.  By building \tool{} on top of AWS
Nitro Enclaves, we inherit their strong security properties; and we carefully
engineered \tool{} to provide seamless and secure networking, scalability, and
remote attestation while remaining entirely user-verifiable.
Our performance evaluation and our
three prototypes suggest that \tool{} can handle low-latency, high-throughput,
and computationally demanding applications like watching HD video streams in an
in-enclave Chromium browser.

\section*{Availability}

Our source code is available online at:\\
\url{https://github.com/brave/nitriding-daemon}

\printbibliography
\balance

\appendix

\section{A basic example}%
\label{sec:example}

Figure~\ref{fig:example} illustrates how an enclave application (called
\texttt{enclave-app}) can run alongside \tool{}.  Figure~\ref{fig:dockerfile}
shows a Dockerfile that adds \tool{}, the enclave application, and a start
script to the image, followed by launching the start script, which is
illustrated in Figure~\ref{fig:start}.  All the script does is first launch
\tool{} in the background followed by launching the enclave application.  If
the application builds reproducibly, it is possible to run it inside an enclave
\emph{without modifications}.

\begin{figure}
  \begin{subfigure}[b]{\linewidth}
    \centering
    \begin{lstlisting}
    FROM alpine:latest

    COPY nitriding /bin/
    COPY enclave-app /bin/
    COPY start.sh /bin/

    CMD ["start.sh"]\end{lstlisting}
    \caption{A Dockerfile that embeds \tool{} along with the enclave
      application, \texttt{enclave-app}.}%
    \label{fig:dockerfile}
  \end{subfigure}

  \begin{subfigure}[b]{\linewidth}
    \centering
    \begin{lstlisting}[language=bash]
    #!/bin/sh

    # Launch nitriding in the background.
    nitriding \
      -fqdn "example.com" \
      -acme \
      -appwebsrv "http://127.0.0.1:8080" &

    # Launch the application.
    enclave-app\end{lstlisting}
    \caption{The start.sh shell script launches \tool{} in the background,
    followed by launching the enclave application}%
    \label{fig:start}
  \end{subfigure}

  \caption{An example of how a simple enclave application can be bundled with
  \tool.}%
  \label{fig:example}
\end{figure}

\section{Architectural diagrams}%
\label{sec:more-diagrams}

\begin{figure}
  \centering
  \begin{tikzpicture}[node distance=20pt] 

  \node [draw,
         label={[anchor=north]above:EC2 host},
         minimum height=90pt,
         align=center,
         minimum width=60pt] (ec2) {};

  \node [draw,
         label={[anchor=north]above:Enclave},
         right=0pt of ec2,
         fill=black!10,
         minimum height=90pt,
         minimum width=60pt] (enclave) {};

  \node [draw,
         below=0pt of enclave.south west,
         minimum width=120pt] (hypervisor) {Hypervisor};

  \node [draw,
         align=center,
         below=of ec2.north] (proxy) {Proxy};

  \node [draw,
         align=center,
         below=of proxy.south] (admin) {Admin};

  \node[draw,
        align=center,
        fill=white,
        xshift=5pt,
        right=of proxy] (nitriding) {Nitriding};

  \node[draw,
        align=center,
        fill=white,
        below=of nitriding] (app) {App};

  \node [draw,
         left=of proxy] (cloudflare) {Cloudflare};

  \node [draw,
         below=of cloudflare] (client) {Client};

  \draw [-latex]
        (admin.north) -- (proxy.south)
        node [midway, fill=white, circle, inner sep=-2pt] {\ding{202}};
  \draw [-latex]
        ([yshift=-5pt]proxy.east) -- ([yshift=-5pt]app.west);

  \draw [-latex, densely dotted]
        (client.east) -- ([yshift=-3pt]proxy.west)
        node [midway, fill=white, circle, inner sep=-2pt] {\ding{203}};
  \draw [-latex, densely dotted]
        ([yshift=5pt]proxy.east) -- ([yshift=5pt]nitriding.west);
  \draw [-latex, densely dotted]
        (nitriding.south) -- (app.north);

  \draw [-latex, densely dashed]
        (app.west) -- (proxy.east)
        node [midway, fill=white, circle, inner sep=-2pt] {\ding{204}};
  \draw [-latex, densely dashed]
        ([yshift=3pt]proxy.west) -- ([yshift=3pt]cloudflare.east);

  \draw [-latex, densely dashdotted]
        (app.south) -- ([xshift=32pt]hypervisor.north)
        node [midway, fill=white, circle, inner sep=-2pt] {\ding{205}};

\end{tikzpicture}
  \caption{An overview of the enclave application that provides verfiable
  configuration transparency.  After launching the enclave, the operator
  configures the confidential bearer token and zone ID~(\ding{202}).  Clients
  can then request the service provider's Cloudflare
  configuration~(\ding{203}).  The application makes an HTTP request (containing
  the bearer token and zone ID) to Cloudflare's API~(\ding{204}).  Finally, the
  application asks its hypervisor for an attestation document~(\ding{205}) and
  embeds the attestation document in the response to the client, along with
  Cloudflare's response.}%
  \label{fig:vct}
\end{figure}
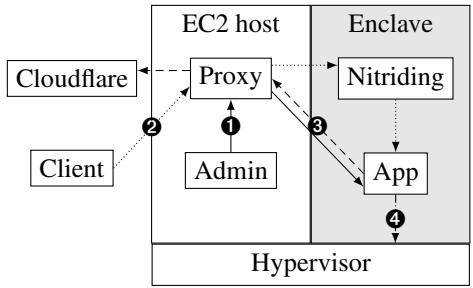

Figure~\ref{fig:vct} illustrates the design of our enclave application which
implements verifiable configuration transparency.

\end{document}